\documentclass[fleqn,usenatbib]{mnras}

\usepackage[T1]{fontenc}
\usepackage{ae,aecompl}
\usepackage{array}
\usepackage{longtable}

\usepackage{graphicx}	
\usepackage{amsmath}	
\usepackage{amssymb}	
\usepackage{enumitem}
\usepackage{nccmath}
\usepackage{cases}

\usepackage{multirow}

\hypersetup{debug}

\newcommand{\upsub}[1]{\sb{\mathrm{#1}}}
\newcommand{\upsup}[1]{\sp{\mathrm{#1}}}

\begingroup\lccode`~=`_\lowercase{\endgroup\let~\upsub}
\begingroup\lccode`~=`^\lowercase{\endgroup\let~\upsup}

\AtBeginDocument{%
  \catcode`_=12 \catcode`^=12
  \mathcode`_="8000
  \mathcode`^="8000
}

\newlist{coloritemize}{itemize}{1}
\setlist[coloritemize]{label=\textcolor{blue}{\textbullet}}

\def\VEL{\textsc{VELOCIraptor}}
\def\TREE{\textsc{TreeFrog}}
\def\WW{\textsc{WhereWolf}}
\def\ORB{\textsc{OrbWeaver}}
\def\SHA{\textsc{Shark}}

\title[Extracting Galaxy Merger Timescales II]{Extracting Galaxy Merger Timescales II: A new fitting formula}

\author[Rhys J. J. Poulton]{
R. J. J. Poulton,$^{1,2}$\thanks{E-mail:  rhys.poulton@icrar.org}
C. Power,$^{1,2}$
A. S. G. Robotham,$^{1,2}$
P. J. Elahi,$^{1,2}$ and
\newauthor \ C. D. P. Lagos$^{1,2}$\\
$^{1}$International Centre for Radio Astronomy Research, University of Western Australia, 35 Stirling Highway, Crawley, WA 6009, Australia \\
$^{2}$ARC Centre of Excellence for All Sky Astrophysics in 3 Dimensions (ASTRO 3D)}

\date{Accepted 15 October 2020}

\pubyear{2020}

\begin{document}
\label{firstpage}
\pagerange{\pageref{firstpage}--\pageref{lastpage}}
\maketitle

\begin{abstract}
Predicting the merger timescale ($\tau_{\rm merge}$) of merging dark matter halos, based on 
their orbital parameters and the structural properties of their hosts, is a fundamental problem 
in gravitational dynamics that has important consequences for our understanding of cosmological 
structure formation and galaxy formation. Previous models predicting 
$\tau_{\rm merge}$ have shown varying 
degrees of success when compared to the results of cosmological $N$-body simulations. We build 
on this previous work and propose a new model for $\tau_{\rm merge}$ that draws on 
insights derived from these simulations. We find that published predictions can provide 
reasonable estimates for $\tau_{\rm merge}$ based on orbital properties at infall, 
but tend to 
underpredict $\tau_{\rm merge}$ inside the host virial radius ($R_{200}$) because tidal stripping 
is neglected, and overpredict it outside $R_{200}$ because the host mass is underestimated. 
Furthermore, we find that models that account for orbital angular momentum via the circular 
radius $R_{circ}$ underpredict (overpredict) $\tau_{\rm merge}$ for bound (unbound) systems. 
By fitting for the dependence of $\tau_{\rm merge}$ on various orbital and host halo properties,
we derive an improved model for $\tau_{\rm merge}$ that can be applied to a merging halo at any point in its orbit. Finally, we discuss briefly the implications of our new model for $\tau_{\rm merge}$ for semi-analytical galaxy formation modelling.
\end{abstract}

\begin{keywords}
methods: numerical - galaxies: evolution - galaxies: haloes
\end{keywords}



\section{Introduction}

The idea that galaxies are embedded in massive, gravitationally-bound, halos of dark matter 
was first proposed in the 1970s \citep[e.g.][]{Freeman1970}, based on observations of the rotation curves of spiral 
galaxies. Subsequent observational evidence  \citep[e.g.][]{Faber1976,Rubin1980} helped to 
establish dark matter as the ``scaffolding of the universe'' \citep[cf.][]{Dayal2018}, tracing 
the distribution of galaxies and the cosmic web, and ultimately this led to the 
emergence of the hierarchical Cold Dark Matter (CDM) model of structure formation. In the CDM 
model, low-mass dark matter halos form via gravitational collapse and merge with other halos 
to grow progressively more massive structures. An integral part of this process is that merging 
halos must lose orbital energy and angular momentum to sink to the centre of the more massive 
host halo before it merges completely. At this point, the merging halo, and any galaxy that
resides in it, is subsumed into the larger halo and is no longer distinct 
\citep{Toomre1977,White1978,Blumenthal1986}. 

Estimating how long this process takes - from the point at which a merging system first
crosses the virial radius of its more massive host, to the point at which it is disrupted and
is no longer a distinct, gravitationally bound, entity - is a theoretical problem of fundamental 
importance. This is most evident in semi-analytical models of galaxy formation and evolution
\citep[SAMs; see][]{baugh_primer_2006,benson_galaxy_2010,somerville_physical_2015}, which 
assume physical prescriptions for galaxy formation that link to the assembly history of
halos parameterised in the form of merger trees, which are most commonly drawn from 
cosmological $N$-body simulations
\citep{Shaun_Cole_et_al_2002,baugh_primer_2006,srisawat_sussing_2013,lee_sussing_2014}. The
timescale for halos and the galaxies they host to merge will impact their predicted properties,
and so an accurate merger timescale is essential. However, finite numerical resolution limits 
the ability of $N$-body simulations to track halos into high overdensity regions over many
orbits and, consequently, to predict the timescale on which merging occurs \citep[e.g.][]{Ostriker1972,Gnedin1999,Dekel2003,Hayashi2003,Kravtsov2004a,Taylor2004,DOnghia2010,VandenBosch2018a,VandenBosch2018}. This has led to the use of the merger timescale ($\tau_{merge}$) 
in SAMs to predict when poorly
resolved halos merge. As a result, it has been explored extensively to understand the
physical properties that govern it, which are formalised in models deduced analytically or from fits to simulation data, and how it impacts 
galaxy mergers  \citep[e.g.][]{Binney1987,lacey_merger_1993,Navarro1995,Velazquez1999,Colpi1999,Jiang2000,Taffoni2003,Zentner2005,Jiang2008,boylan-kolchin_dynamical_2008,gan_modelling_2010,Mo2010,Wetzel2010a,jiang_n-body_2014,Simha2017}. 

\medskip

The merger timescale was first formulated analytically by \citet{Binney1987}\footnote{\label{fn:note1}We 
note that both the \citet{Binney1987} and \citet{lacey_merger_1993} models are not formally for 
$\tau_{merge}$ because their focus is dynamical friction only, and not the variety of processes 
that are present in an $N$-body simulation. However, we include them in our work because they 
have been used as a proxy for $\tau_{merge}$ in some SAMs.} (BT87 hereafter), which is based on 
Chandrasekhar's dynamical friction prescription \citep{Chandrasekhar1943} -- the primary (but not 
sole) mechanism by which halos lose their orbital energy \citep{Jiang2008,boylan-kolchin_dynamical_2008}. This formulation assumes that the infalling
halo (or subhalo once it crosses the host virial radius $R_{vir,host}$) is a point mass, moving at velocities much less than the local velocity dispersion of the host through a uniform background of low-mass point masses. Subsequently, 
\citet{lacey_merger_1993}\footnotemark[1] (LC93 hereafter) proposed an analytical model that treats the 
host halo as a singular isothermal sphere and integrates the orbit-averaged equations to find the 
$\tau_{merge}$ based on the initial orbital energy and orbital angular momentum. This inclusion of
energy and angular momentum, through use of the circularity ($\eta$) and circular radius ($R_{circ}$)\footnote{Recall that the circularity is defined as, $\eta=J_{\rm halo}/J_{\rm circ}(E)$, where $J_{\rm halo}$ is the specific angular momentum of the orbiting (sub)halo and $J_{\rm circ}(E)$ is the specific angular momentum of the equivalent circular orbit with the same energy ($E$). The corresponding
circular radius is $R_{\rm circ}=GM_{\rm encl,host}(r)M_{\rm sub}/2E$, where $M_{\rm sat}$ is the subhalo mass and $M_{\rm encl,host}(r)$ is the enclosed mass at
the radius of the subhalo.},means that this formulation can account for varying orbits.

Following on from the analytical work of BT87 and LC93, \citet{Jiang2008} (hereafter J08)
and \citet{boylan-kolchin_dynamical_2008} (hereafter BK08) used numerical simulations to 
refine models for $\tau_{merge}$. J08 investigated the performance of the LC93 
formulation by comparing its predictions with hydrodynamical $N$-body simulations, and found that 
it systematically underpredicts the timescale for minor mergers and overpredicts it for major 
mergers, which they distinguish by a 3:1 mass ratio. This led them to propose a new model for 
$\tau_{merge}$ by fitting to the simulations as a function of $\eta$ and $R_{circ}$. At the same 
time, BK08 use idealised simulations of a host halo and merging satellite, run with 
different mass ratios, $\eta$ and $R_{circ}$, to assess how orbital energy and angular 
momentum modifies $\tau_{merge}$ as formulated by BT87. This led them to obtain a fit to
$\tau_{merge}$ from their simulations by parameterising the dependence on host-to-satellite mass 
ratio, $\eta$ and $R_{circ}$.  

\medskip

In the preceding paper in this series \citep{Poulton2019}, we used accurate merger trees
to characterise the orbital properties of the (sub)halo population in cosmological $N$-body 
simulations, and assessed the performance of the models proposed by BT87, LC93, J08, and BK08
for predicting $\tau_{merge}$. We concluded that all the considered models provide a reasonable 
approximation for systems with short-to-intermediate $\tau_{merge}$, but they tend to
overpredict $\tau_{merge}$ for the smallest mass systems. Moreover, the values for 
$\tau_{merge}$ that are based on models derived from simulations (J08 and BK08) perform well when predicting 
$\tau_{merge}$ for subhalos whose properties are comparable to those used in the
original studies - principally set by the number of particles in a subhalo and its
mass ratio relative to its host - but tend overpredict it by a factor of 100 when we consider subhalos with higher mass ratios, probing a regime that could not be resolved
in the original studies.

In this paper, we investigate the physical factors that underpin the various models for $\tau_{merge}$ 
to understand what drives their performance in the various regimes, and what gives rise to the
discrepant behaviours highlighted in \citet{Poulton2019}. We propose a new, more accurate, 
model for $\tau_{merge}$ that includes the effects of dynamical friction, dynamical 
self-friction \citep{Miller2020}, tidal stripping, and tidal heating. This new model is applicable 
to a large dynamic range of subhalo or satellite mass to host mass ratios, in a large variety of environments, 
and can be applied at any point in a subhalo's or satellite's orbit around its host.

The remainder of this paper is organised as follows: in \S2, we describe the input simulation, 
the calculations used, and the sample of subhalos and hosts that we use in our analysis. In \S3, we discuss the current 
$\tau_{merge}$ predictions; we present our new model; and we assess its performance compared 
to the current prescriptions.  Finally, in \S4, we discuss our results, 
and we present our  conclusions and their implications. Throughout this paper, we use virial quantities for the host 
halo, which are defined using 200$\rho_{crit}$, where $\rho_{crit}$ is the critical density of 
the universe\footnote{We could use other definitions of host halo mass and radius, which would 
change some parameters in our model.}. We assume a $\Lambda$CDM cosmology in 
accordance with the Planck Collaboration data \citep{planck_collaboration_planck_2015} - with density parameters $\Omega_{M}$ = 0.3121, $\Omega_{b}$ = 0.6879, and $\Omega_{\Lambda}$ = 0.6879; a normalisation $\sigma_{8}$  = 0.815; a primordial spectral index $n_{s}$ = 0.9653; and ${\rm H}_0=67.51 { \rm km\,s}^{-1}{\rm Mpc}^{-1}$. Note that we use the terms subhalo and satellite interchangeably in this paper.

\section{Methods}

\subsection{Simulations \& Halo Tracking}

The simulations used in this work come from the \textsc{genesis} suite of $N$-body simulations, with volumes ranging from 26.25 to 500 Mpc $h^{-1}$ and between 324$^3$ to 5200$^3$ particles. We focus on the 105 $h^{-1} {\rm Mpc}$ box with $2048^3$ particles simulation because it allows us to probe well-resolved hosts ranging in mass from 10$^{12}$ to 10$^{14}M_{\odot}$. This simulation has a total of 190 snapshots, evenly spaced in logarithmic expansion factor ($a=1/(1+z)$) between $z = 24$ to $z = 0$. This high cadence (on average 70 Myrs between snapshots) enables an accurate capturing of the evolution of dark matter halos and their orbits.

Halo catalogues are constructed using \VEL, a 6-Dimensional Friends-of-Friends (6D-FoF) phase space halo finder \citep{elahi_peaks_2011,elahi_streams_2013,Poulton2018,Canas2019,Elahi2019VELOCIraptor}, while trees are constructed using \TREE\ \citep{Poulton2018,Elahi2019TreeFrog}, which is a particle correlator that can link across multiple snapshots and halo catalogues. Importantly, \TREE's ability to link across multiple snapshots is vital for tracking subhalos as they orbit within highly overdense regions. While a subhalo may not be present in a pair of halo catalogues at consecutive output times, it may be present in halo catalogues at a later time, and so there might be gaps in the subhalo's history. This has led to the development of the halo tracking tool known as \WW, as described in \citep{Poulton2018,Poulton2019}. \WW\ is a halo ghosting tool, designed to fill in the gaps in a subhalo's history by using the particles from the simulation and attempting to find the missing halo by using the subhalo's bound particles and calculating its properties. \WW\ also attempts to extrapolate when the halo merges, so that subhalos are tracked until they coalesce with their host.

To extract all subhalos relative to their host and calculate their orbital properties, we use \ORB\ as described in \citet{Poulton2019}. \ORB\ first identifies host halos of interest;
identifies (sub)halos that come within N $\times$ R$_{ vir,host}$; extracts the full history of the (sub)halo relative to its host; identifies key points in its orbit; before finally 
interpolating the orbit to obtain characteristic properties of the orbit at these key points. Using \textsc{OrbWeaver} we calculate $\tau_{merge}$ as being the exact time from infall (crossing $R_{vir,host}$) to the time when the halo is no longer a self-bound entity (has phase mixed with its host).

\subsection{Calculating orbital properties}
\label{ssec:orb_props}
We calculate halo/subhalo positions and velocities (in physical coordinates and including the Hubble flow) relative to the centre of potential of the host.  We compute halo/subhalo orbital properties by treating them as isolated point particles in the reduced mass frame, while separations and relative velocities with respect to the
host are calculated as  $\textbf{r} = \textbf{r}_{host} - \textbf{r}_{sat}$ ($\textbf{v} = \textbf{v}_{host} - \textbf{v}_{sat}$). The reduced mass is  $\mu = M_{ sat}M_{ encl,host}(r)/(M_{ sat} + M_{ encl,host}(r ))$, where  $M_{ sat}$ is the mass of the subhalo and $M_{encl,host}(r )$ is the enclosed mass of the host out to radius $r$, which is calculated assuming a Navarro, Frenk and White (NFW) profile \citep{navarro_universal_1997},
with a characteristic density\footnote{This is a property of the halo.} given by,

\begin{ceqn}
\begin{equation}
\rho_{0}  = \frac{M_{vir,host}}{4 \pi R_{s}^3 \left[ \mathrm{ln} \left(1 + c\right) - \left(\frac{c}{1 + c}  \right) \right]}.
\label{equ:rho}
\end{equation}
\end{ceqn}

\noindent
Here c is the NFW concentration parameter $c=\frac{R_{vir}}{R_{s}}$. M$_{encl,host}$ is given by

\begin{equation}
M_{encl,host}(r)=
\begin{cases}
4 \pi \rho_{0} R_{s}^3 \left[ \mathrm{ln} \left(1+\frac{r}{R_{s}}\right) - \left(\frac{r}{R_{s} + r}  \right) \right] \ (r < R_{vir,host}) \\
\\
 M_{vir,host} \hspace{3.6cm} (r \geq R_{vir,host}).
\end{cases}
\label{equ:Mencl}
\end{equation}

\noindent
The orbital energy of the subhalo\footnote{We note that the following equation is for a point mass, but it provides a good approximation and is a efficient means of calculating the orbital state.} is given by:
\begin{ceqn}
\begin{equation}
E = 0.5 \mu \mathrm{v}^{2} \  - \  \frac{GM_{encl,host}(r)M_{sat}}{r},
\label{equ:E}
\end{equation}
\end{ceqn}

\noindent
and its orbital angular momentum is 
\begin{ceqn}
\begin{equation}
L = \mu \textbf{r}  \times  \textbf{v}.
\label{equ:L}
\end{equation}
\end{ceqn}
\noindent
The eccentricity of the orbit can be computed from
\begin{ceqn}
\begin{equation}
e = \sqrt{1 + \frac{2EL^2}{(GM_{encl,host}(r)M_{sat})^2 \mu}},
\end{equation}
\end{ceqn}
and the pericentric distance from the host centre is given by
\begin{ceqn}
\begin{equation}
R_{peri} = \frac{L^2}{(1 + e)GM_{encl,host}(r)M_{sat} \mu}.
\label{equ:rperi}
\end{equation}
\end{ceqn}

\noindent
It is also possible to compute the dynamical time of a subhalo at radius r assuming: 
\begin{ceqn}
\begin{equation}
T_{dyn}(r) =  \sqrt{\frac{r^3}{GM_{encl,host}(r)}}.
\label{equ:Tdyn}
\end{equation}
\end{ceqn}

\subsection{Merging sample} \label{sec:samplesel}

\subsubsection{Selection criteria}

Our goal is to create an analytical formulation for $\tau_{merge}$ for objects that merge with their host\footnote{We consider all merger events up to $z$ = 0 to maximize statistics.}, but to ensure that we have objects that can be treated (to first order) as members of two-body systems (so equation \ref{equ:rperi} can be applied), we apply the same selection as presented \citet{Poulton2019} to select our hosts:

\begin{description}[align=left,leftmargin=0.5cm,style=nextline]
\item[Top of its spatial hierarchy:] 
This excludes satellites orbiting satellites, which make up a negligible number of the orbiting sample.
\item[N$_{host}$> 10,000 particles:] 
The host potential must be well sampled so that orbital histories for the orbiting systems can be accurately recovered.
\end{description}

\noindent These selection criteria guarantee that our host selection contains well-resolved field halos and removes any satellite-satellite mergers (which represent only 10\% of mergers in a typical simulation; see \citealt{Deason2014}).

\smallskip

Ideally, we would like to track satellites until they coalesce physically with their hosts, but finite time sampling and insufficient numerical resolution means that satellites can disrupt and coalesce unphysically with their host before this point. To circumvent these issues, we first select systems that satisfy (see Appendix \ref{app:Sel} for how these were selected):

\begin{description}[align=left,leftmargin=0.5cm,style=nextline]
\item [Merge within 0.1 R$_{ vir,host}$ ] 
A satellite must be within 0.1 R$_{ vir,host}$ when it is lost, and be a secondary progenitor of its host. 
\item[N$_{sat}$ > 1000 particles at infall ] Halos that are satellite progenitors cannot disrupt before they come within R$_{ vir,host}$ \citep[cf.][]{VandenBosch2018}.
\end{description}

\smallskip
\noindent Finally, we require that systems must satisfy at all times:

\begin{description}[align=left,leftmargin=0.5cm,style=nextline]
\item [N$_{sat}$ >100 particles:] 
The mass of the satellite must be well above the threshold for detection by the halo finder.
\item [R$_{ s}$ > 2.0 $\times$ softening length:]  
The satellite is sufficiently well resolved in the simulation such that it is not susceptible to artificial disruption due to force softening.
\end{description}

\noindent These criteria result in 8930 mergers in our simulation that we can use in our analysis. We use every point between 0.1 to 3.0 R$_{ vir, host}$ in the merging system's orbital history. 

\begin{figure}
    \centering
    \includegraphics[width=0.49\textwidth]{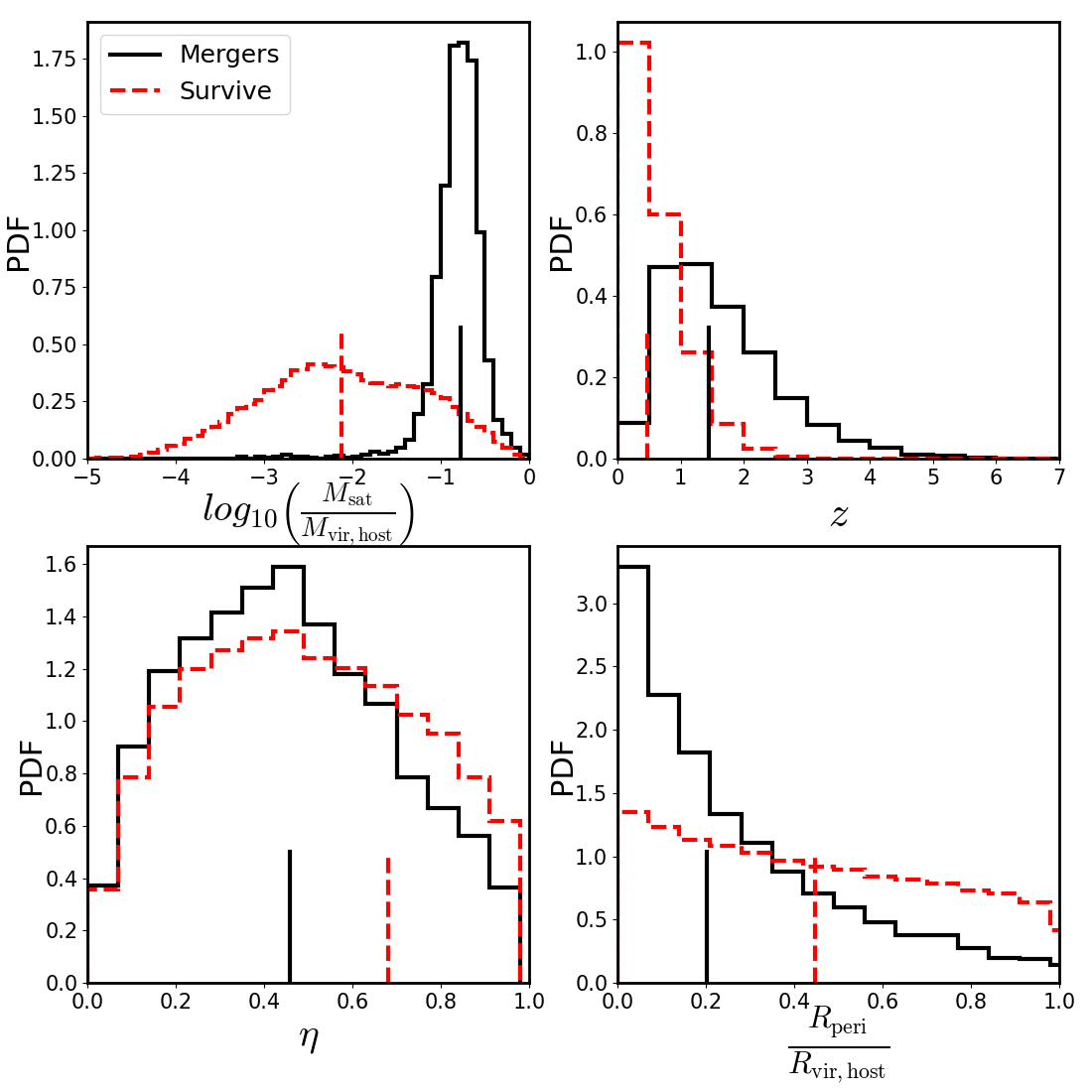}
    \caption{The basic statistical properties of the merging sample calculated at infall. \textit{Top left}: the mass ratio between the satellite and the host. \textit{Top right}: the redshifts where these objects are found. \textit{Bottom left}: circularity $\eta$. \textit{Bottom right}: the ratio of R$_{ peri}$ from equation \ref{equ:rperi} to R$_{ vir,host}$. The solid black line shows the distribution for the merging sample and the dashed red line shows the sample that survives until $z$=0. The vertical black and dashed red lines shows the medians of the distributions.}
    \label{fig:dist}
\end{figure}

\subsubsection{Sample distributions}

Figure \ref{fig:dist} shows the basic statistical properties of the merging sample, as well as 
for all systems that survive until $z$=0 in our selected hosts. In our merging sample, there are 
2 times more minor mergers than major mergers, where we split major to minor at a 3:1 mass ratio. 
Systems that merge tend to have mass ratios closer to 1:1 than systems that survive. The surviving
sample falls onto the host later than the merging sample, with median infall redshifts of $z_{\rm infall} \simeq $ 1.5
for the merging sample and $z_{\rm infall} \simeq $ 0.5 for the surviving sample. This difference in median infall redshift
indicates that it is interaction with the host environment rather than intrinsic properties of the
infalling systems that drives merging.

The merging sample tends to be on more radial orbits compared to the surviving sample, with 
a median of 0.45 for merging systems and 0.7 for surviving systems, and they also tend to have 
smaller values of $R_{ peri}$ / $R_{vir,host}$, with a median of 0.2 compared to 0.45 for the 
surviving sample. This difference between merging and surviving samples suggests the predictive 
capability of $R_{peri}$ for determining mergers in the time of the simulation. 

\section{A New Model for the Merger Timescale}

In this section, we evaluate how commonly used published formulae (cf. Table \ref{tab:form}) for $\tau_{\rm merge}$ - from BT87, LC93, J08, BK08 - perform as a function of $r$ / $R_{vir,host}$, and present our new 
formulation for $\tau_{\rm merge}$ and its variation with $r$ / $R_{vir,host}$. We also discuss the dependence of $\tau_{\rm merge}$ on orbital energy.

\subsection{Performance of published models}

\begin{figure}
    \centering
    \includegraphics[width=0.49\textwidth]{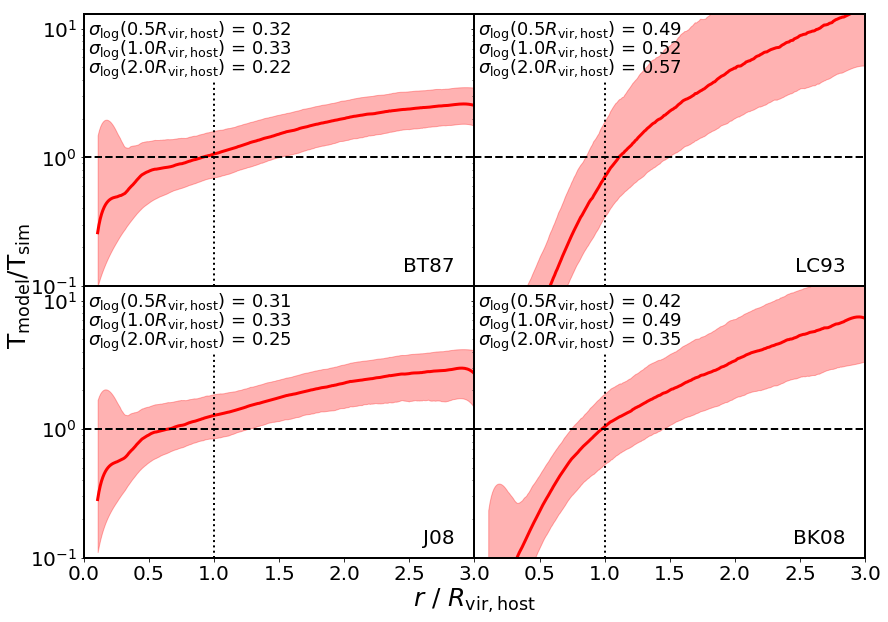}
    \caption{The ratio of the existing $\tau_{merge}$ prescriptions predictions (T$_{model}$) to the measured $\tau_{merge}$ from the simulation (T$_{sim}$) change as a function of $r$ / $R_{vir,host}$. The text in the bottom right of each panel states the T$_{model}$ that is shown. The red line shows the median of the population and the shaded region show the standard deviation. The text in the top left of each panel states the standard deviation in T$_{model}$/T$_{sim}$ at 0.5, 1.0, and 2.0 $R_{ vir,host}$.}
    \label{fig:MTSVSR}
\end{figure}

\begin{table*}
\caption{The published $\tau_{merge}$ prescriptions formulations. The left most column is the $\tau_{merge}$ prescription, the middle column is the parameter values for each prescription and the right most column is the standardised formula used by all these prescriptions.  \label{tab:form}}
\resizebox{\textwidth}{!}{
\begin{tabular}{cccccccc}
\hline 
Model & \multicolumn{5}{c}{Parameters} & $\tau_{merge}$ formula \\ \hline \\[-8pt]
BT87  & A=1.17 & $\Lambda$ = 1 + $\frac{M_{host}}{M_{sat}}$ & b=1.0 & c=0 & $f(\eta)$=1 & \multirow{4}{*}{$A\left(\frac{T_{dyn}}{\mathrm{ln}(\Lambda)}\right) \left(\frac{M_{host}}{M_{sat}}\right)^{b} \left(\frac{R_{circ}}{R_{vir,host}}\right)^{c} f(\eta)$}
\\[2pt] 
LC93 & A=1.17 & $\Lambda$ = $\frac{M_{host}}{M_{sat}}$ &  b=1.0 & c=2 & $f(\eta)$=$\eta^{0.78}$  &
\\[2pt] 
BK08 & A=0.216 & $\Lambda$ = 1 + $\frac{M_{host}}{M_{sat}}$ & b=1.3 & c=1 & $f(\eta)$=$\mathrm{exp}(1.9\eta)$ &
\\[2pt] 
J08 & A=1.17 & $\Lambda$ = 1 + $\frac{M_{host}}{M_{sat}}$ & b=1.0  & c=0 & $f(\eta)$=0.94$\eta^{0.6}$ + 0.6 &
\\[5pt] \hline 
\end{tabular}
}
\end{table*}

We compute predictions for $\tau_{merge}$ for merging systems at all points along their orbits and
compare with $\tau_{merge}$ measured directly from the orbital history deduced from the merger
tree data. In practice, we evaluate the ratio, T$_{model}$/T$_{sim}$, of the timescale predicted by the model, T$_{model}$, and the measured simulation timescale, T$_{sim}$, and show how this varies with $r/R_{vir,host}$ in Figure \ref{fig:MTSVSR}. 

\subsubsection{\citet{Binney1987} (BT87)}
We find that the BT87 model recovers T$_{model}$/T$_{sim}$ $\simeq$ 1 when evaluated at 
$r = R_{vir,host}$, but tends to underpredict $\tau_{merge}$ for $r < R_{vir,host}$ and 
overpredict it at $r > R_{vir,host}$. The underprediction at smaller radii arises because the 
model neglects the effects of tidal stripping, which can prolong the lifetime of a satellite 
\citep{Jiang2008,boylan-kolchin_dynamical_2008,Mo2010}, whereas the overprediction at larger 
radii is a consequence of the assumption that accreting halos are on smoothly inspiralling 
orbits that do not change as the system falls onto its host. However, this assumption does not 
hold in practice; most orbits transition from being preferentially radial at larger radii to
isotropic within the host (see appendix \ref{app:CircVsR}), and so accreting halos at larger 
radii take a shorter time to cross the region $1 < r/R_{vir,host} < 3$ than predicted by BT87.

\subsubsection{\citet{lacey_merger_1993} (LC93)}
The LC93 model shows a strong dependence on $r$/$R_{vir,host}$, which reflects its dependence on
$R_{circ}$ and $\eta$ and on the assumption that the virial theorem is valid ($U=2T$). However, this 
assumption breaks down when considering systems at the $R_{vir,host}$ (see appendix \ref{app:virial} 
for more information) and leads to the large change with $r$ / $R_{vir,host}$. The dependence on 
$r$ / $R_{vir,host}$ is exacerbated by the neglect of tidal stripping. 
Furthermore, the LC93 formula has an offset such that T$_{model}$=T$_{sim}$ at $r$ = 1.1$R_{vir,host}$ rather than 
$r$ =$R_{vir,host}$, which is most likely due to the assumption that halos are isothermal spheres, which leads them to have larger $R_{vir,host}$. The LC93 model shows the largest scatter of the models, principally because of the inclusion of $R_{circ}$ and $\eta$ in its
formulation; the scatter is driven by the large range of boundedness of orbits (shown in 
appendix \ref{app:virial}).

\subsubsection{\citet{Jiang2008} (J08)}
The J08 model has the same functional form as the BT87 model, and so it suffers from the same 
behaviours at small and large radii, with similar scatter. Although J08 calibrated their model
using a simulation that naturally accounted for the effects of tidal stripping, the halos and
satellites in their sample covered a restricted mass range, with $M_{sat}$/$M_{vir,host}>0.1$.
This means that their merging systems had smaller $\tau_{merge}$ and so they did not survive
sufficiently long to experience the full effects of tidal stripping. Interestingly, the J08 
formula predicts T$_{model}$=T$_{sim}$ at $r$ = 0.9$R_{vir,host}$ rather than 
$r$ =$R_{vir,host}$, which may reflect the use of a hydrodynamical simulation for 
calibration. This would be consistent with the findings of previous studies, which 
showed that the inclusion of baryons reduces $\tau_{merge}$ for merging satellites by 10\%, 
\citep{boylan-kolchin_dynamical_2008,Jiang2010}. We note that J08 based their results solely
on the orbital properties of infalling halos and satellites at $r$=$R_{vir,host}$, and so
did not have information about how their formulation for $\tau_{merge}$ varied with radius.

\subsubsection{\citet{boylan-kolchin_dynamical_2008} (BK08)}

The BK08 model is similar to the LC93 model insofar as it accounts for both energy and angular 
momentum ($\eta$, $R_{circ}(E)$) in its formulation. Consequently, it also suffers from the same 
issues as the LC93 model, albeit to a lesser extent. In common with J08, BK08 based their 
results solely on the orbital properties of infalling halos and satellites at 
$r$=$R_{vir,host}$. Furthermore, they calibrated their model against idealised simulations
of merging satellites with their host halos, and so they did not probe the full range of orbits present in 
cosmological simulations.

\subsection{A new model}

In constructing the new model for $\tau_{merge}$, we use the hyperplane fitting package, \textsc{hyper.fit} \citep{Robotham2015}, which uses a likelihood analysis to minimise the scatter in the hyperplane. This allows us to fit a parameterised functional form to the data and to find the key predictive quantities that minimise the scatter in T$_{model}$/T$_{sim}$. 

\smallskip

\noindent  Our choice of functional form is, 
\begin{ceqn}
\begin{equation}
\mathrm{T}_{fit} = A \ T_{dyn}(r) \left(\frac{r}{R_{vir,host}}\right)^b \left(\frac{R_{peri}}{R_{vir,host}}\right)^c.
\label{equ:fit}
\end{equation}
\end{ceqn}
This predicts orbits that agree well with the data (see Appendix \ref{app:orbpred}) and it accounts for the expected dependence of $\tau_{merge}$ on a satellite's 
position within its host and the dynamical time of the host. The various physical quantities are as defined in section \ref{ssec:orb_props}, and $A$, $b$, and $c$ are free parameters. With this definition, we favour values of  $A$ = 5.5 and 
$c$ = 0.2, while the constant $b$ is dependent on the satellite's position - specifically, whether it is inside or outside R$_{ vir,host}$\footnote{The extra radial dependence is required inside $R_{vir,host}$ due to the use of $M_{ encl,host}(r)$ that adds an extra radial dependence, which is not present outside $R_{vir,host}$ because $M_{vir,host}$ is used.}. We find, 
\begin{ceqn}
\begin{equation}
\begin{aligned}
    b = -0.5 && (r < R_{vir,host}), \\ 
    b = -1.0 &&  (r \geq R_{vir,host}),
\end{aligned}
\end{equation}
\end{ceqn}
which means that there is a stronger dependence on position when outside the host. Equation~\ref{equ:fit} simplifies to:

\begin{numcases}{\mathrm{T}_{fit}=}
5.5 \sqrt{\frac{R_{ vir,host}}{GM_{ encl,host}(r)}} r^{0.8} R_{peri}^{0.2} \hspace{0.2cm}  (r < R_{vir,host}), \label{equ:Tnewin}  
\\
5.5 \frac{R_{vir,host}}{\sqrt{G M_{vir,host}}} r^{0.3} R_{peri}^{0.2} \hspace{0.75cm} (r \geq R_{vir,host}), \label{equ:Tnewout}
\end{numcases}

which most compactly captures the behaviours at $r < R_{vir,host}$ and $r \geq R_{vir,host}$.

\subsubsection{Performance}

\begin{figure}
    \centering
    \includegraphics[width=0.49\textwidth]{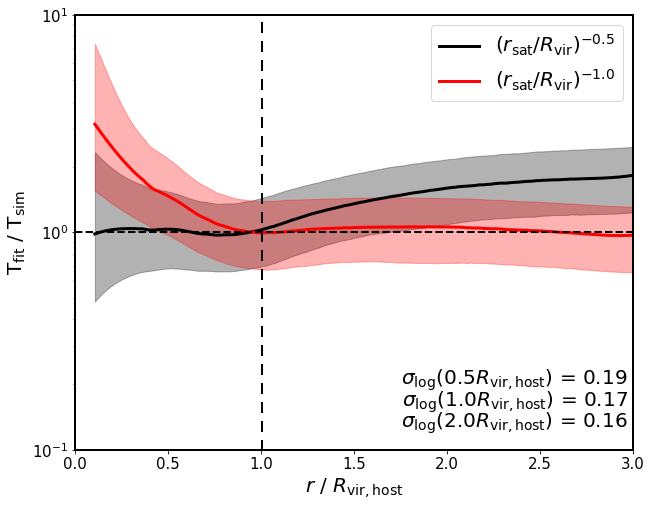}
    \caption{Variation of T$_{fit}$/T$_{sim}$ with $r$/$R_{vir,host}$. Solid lines show medians of the population, whike the shaded region shows the standard deviation. Different colours show the different dependencies of Equation
    \ref{equ:Tnewout} on $r$ / $R_{vir,host}$, shown in the legend.}
    \label{fig:tnew}
\end{figure}

Figure \ref{fig:tnew} shows the performance\footnote{We also demonstrate the performance of the the new model when applied in a lower resolution simulation in Appendix \ref{app:conv}.} of the new model as a function of $r/R_{vir,host}$. Equation \ref{equ:Tnewin} performs well at $r < R_{vir,host}$ but degrades at $r > R_{vir,host}$, while Equation \ref{equ:Tnewout} performs well at $r > R_{vir,host}$ but degrades at $r < R_{vir,host}$. This motivates the need for two overlapping formulae to characterise the 
behaviour of T$_{fit}$/T$_{sim}$ over the radial range of interest.

\begin{figure}
    \centering
    \includegraphics[width=0.49\textwidth]{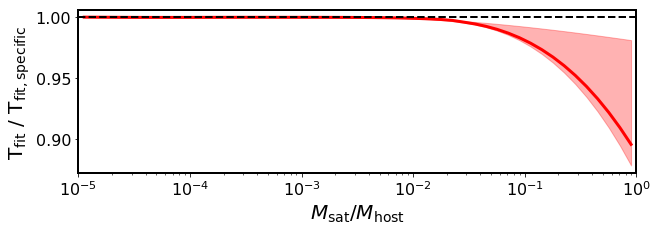}
    \caption{Variation of the ratio of T$_{fit}$ to the specific $\tau_{merge}$, $T_{new,specific}$, with the ratio $M_{sat}$ / $M_{host}$. The red line shows the median of the population and the shaded region shows the standard deviation.}
    \label{fig:tnewnomsat}
\end{figure}

We note that the new model has limited dependence on the satellite-to-host mass ratio, such that the dependence on $M_{sat}$\footnote{This mass corresponds to both bound and unbound particles, and correspond to particles grouped together by the 6D-FoF algorithm at each simulation output. The mass is the mass of the satellite at the point where the formula is being evaluated.} is driven by the calculation of $R_{peri}$ from Equation \ref{equ:rperi}. However, this dependence on $M_{sat}$ can be removed completely by using the specific energy in Equation \ref{equ:E} and specific angular momentum in Equation \ref{equ:L}. Removing this dependence on $M_{sat}$ is useful because the mass of the satellite can depend upon the choice of halo-finder used, particularly for objects deep inside their host potential \citep{knebe_haloes_2011,muldrew_measures_2012,Poulton2018}.

Removing $M_{sat}$ from equations \ref{equ:Tnewin} and \ref{equ:Tnewout} yields the specific $\tau_{merge}$ ($T_{new,spec}$), given by:
\begin{equation}
\mathrm{T}_{fit,spec} =
\begin{cases}
     5.5 \ \sqrt{\frac{R_{ vir,host}}{GM_{ encl,host}(r)}} r^{0.8} R_{peri,spec}^{0.2}  &  (r < R_{vir,host}), \\
     \\
   5.5 \ \frac{R_{vir,host}}{\sqrt{G M_{vir,host}}} r^{0.3} R_{peri,spec}^{0.2} &  (r \geq R_{vir,host}), 
\end{cases}
\label{equ:Tnewspec}
\end{equation}
where $R_{peri,spec}$ is:
\begin{ceqn}
\begin{equation}
R_{peri,spec} = \frac{l^2}{(1 + e_{spec})GM_{encl,host}(r) },
\label{equ:rperispec}
\end{equation}
\end{ceqn}
where $l$ is the specific angular momentum and $e_{spec}$ is:
\begin{ceqn}
\begin{equation}
e_{spec} = \sqrt{1 + \frac{2 \varepsilon l^2}{(GM_{encl,host}(r)^2} },
\end{equation}
\end{ceqn}
where $\varepsilon$ is the specific orbital energy.

To show the effects of using the specific angular momentum and energy, we plot the ratio of T$_{fit}$ to $T_{new,spec}$ as a function
of $M_{sat}$ / $M_{host}$. Figure \ref{fig:tnewnomsat} shows that for most mass ratios that there is good agreement between T$_{fit}$ and $T_{new,spec}$ and only starts to over predict $\tau_{merge}$ when the  is a tenth of its host mass. The difference is about 10\% for the most massive satellites, which is smaller than the differences in the difference in the recovered mass by different halo finders \citep{knebe_haloes_2011}.

\subsection{Dependence on orbital energy}

\begin{figure*}
    \centering
    \includegraphics[width=0.95\textwidth]{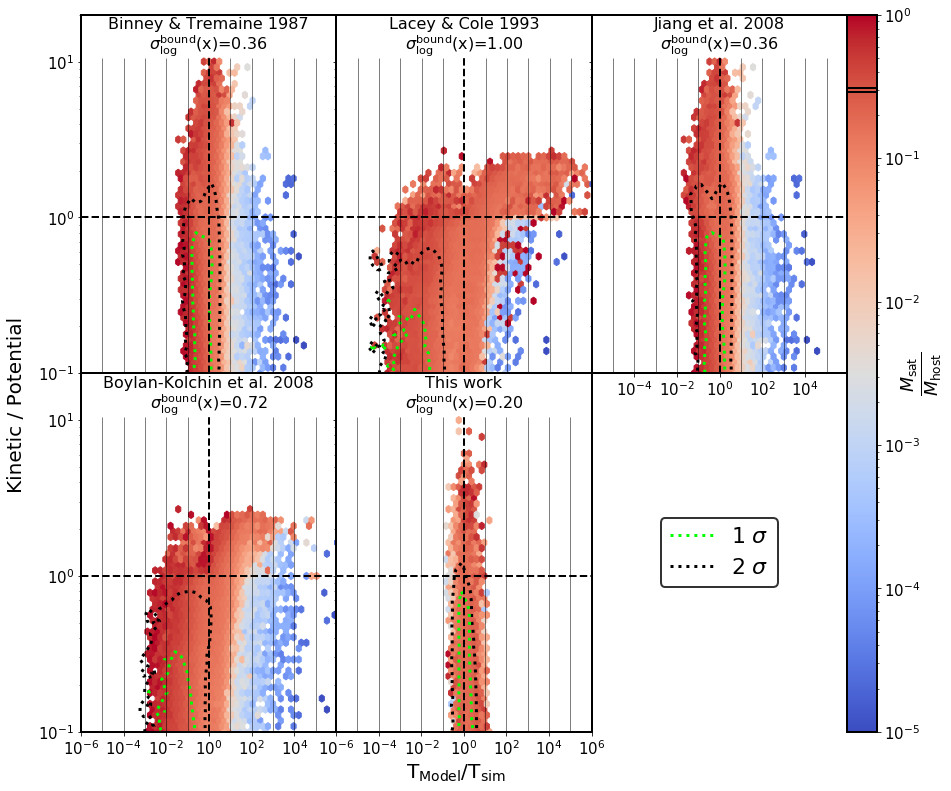}
    \caption{2D histograms of the joint distribution of the ratio of T$_{model}$ over T$_{sim}$ and the
    ratio of orbital kinetic energy (Kin) to potential energy relative to its host (Pot). The text in the top of each panel states the T$_{model}$ that is shown and the standard deviation of the log$_{10}($T$_{model}/$T$_{sim})$ for the bound systems in the panel. The horizontal dashed line shows the region which is 
    unbound above and bound below. The vertical dashed line shows where T$_{model}$=T$_{sim}$. The dotted green and black lines on each panel show the 1$\sigma$ and 2$\sigma$ contours, respectively. The colour shows the median satellite-to-host mass ratio in each bin and the box on the colour bar highlights the colour for the median mass of satellites, given no variation in mass \label{fig:MTSRVsE}}
    
\end{figure*}

Figure \ref{fig:MTSRVsE} shows a 2D histogram of the ratio of $T_{model}$/$T_{sim}$ plotted against the 
ratio of orbital kinetic energy (KE) to potential energy relative to its host (GPE),
with colour coding accounting for the satellite-to-host mass ratio. This Figure shows how the 
different models perform, even when satellites are currently unbound from their host (i.e. KE > 
GPE). All of the currently published models (top panels, bottom left panel) tend to 
underpredict $\tau_{merge}$ for larger satellite-to-host mass ratios, and overpredict for 
smaller mass ratios, reflecting the strong dependence on satellite-to-host mass ratio in the 
model functional forms.

The Figure also shows that both the LC93 and BK08 models have a dependence on the KE to GPE 
ratio - when satellites become unbound from the host, T$_{sim}$ tends to be overpredicted. This 
reflects the dependence of these models on $R_{circ}$, which becomes large when KE $\sim$ GPE 
(shown in Appendix \ref{app:virial}). It's also noteworthy that the LC93 model overpredicts the 
lifetimes of the largest systems (i.e. $M_{sat}$/$M_{host} \approx$ 1) even when they are bound, which arises because the Coulomb logarithm
$\log(\Lambda)\rightarrow\infty$ as $M_{sat}/M_{host}\rightarrow 1$.

Our new model (bottom right panel) performs well for all masses and for all types of orbit. 
As for the BT87 and J08 models, our model's predictions do not change when the satellite becomes 
unbound. Furthermore, T$_{fit}$ does not show the residual dependence on the satellite-to-host 
mass shown that was evident in the currently published models. The new model shows reduced 
scatter, even for the most unbound satellites, which is because it uses $R_{peri}$ to capture
the dependence on satellite orbit. We discuss why $R_{peri}$ is a better predictor of orbit than 
$\eta$ in Appendix \ref{app:orbpred}.

\section{Conclusions}

We have investigated the performance of the most commonly used, published, models that predict the timescale 
for merging ($\tau_{merge}$) between a satellite and its more massive host (\citet{Binney1987} (BK08); \citet{lacey_merger_1993} (LC93); \citet{Jiang2008} (J08), and \citet{boylan-kolchin_dynamical_2008} (BK08)) and used the results of this analysis to motivate a new model based on our own cosmological $N$-body simulations. Guided
by the behaviour of $\tau_{merge}$ predicted by the models inside and outside of the
more massive host's virial radius, we favour a dual formulation for $\tau_{merge} \equiv T_{fit}$,
such that its calculation depends on the value of $r/R_{vir,host}$,
\begin{numcases}{\mathrm{T}_{fit}=}
5.5 \sqrt{\frac{R_{ vir,host}}{GM_{ encl,host}(r)}} r^{0.8} R_{peri}^{0.2} \hspace{0.2cm}  (r < R_{vir,host}), \nonumber
\\
5.5 \frac{R_{vir,host}}{\sqrt{G M_{vir,host}}} r^{0.3} R_{peri}^{0.2} \hspace{0.75cm} (r \geq R_{vir,host}). \nonumber
\end{numcases}

This new model accurately estimates $\tau_{merge}$ for all infalling halos and satellites within 3$R_{vir,host}$. 

Based on our analysis, we have drawn the following conclusions:
\begin{itemize}
    \item All currently published models show a strong dependence on $r$/$R_{vir,host}$. For the BT87 and LC93 models, this is because they do not take into account tidal stripping and, in the case of the BT87 model, do not include the effect of the type of orbit the merging system is on. The J08 and BK08 models were calibrated at $r=R_{vir,host}$, and so they did not account for the trend with $r$/$R_{vir,host}$. 
    \item The predictions of the LC93 and BK08 models depend on the circular radius $R_{circ}$, which affects not only their median behaviour with radius, but also the size of scatter at fixed radius, which is a consequence of the dependence of $R_{circ}\equiv R_{circ}(E)$ on orbital energy, $E$. 
    \item This dependence on orbital energy means that both the LC93 and BK08 models overpredict 
    $T_{sim}$ for satellites and halos on unbound orbits; this is not the case for the BT87 or
    J08 models, or our new model.
    \item All currently published models predict $\tau_{merge}$ that show a strong dependence on satellite-to-host mass ratio, underpredicting $\tau_{merge}$ for higher mass systems and overpredicting it for lower mass ones. Our prescription does not show this same behaviour and provides an accurate prediction for all satellite-to-host mass ratios.
    \item The pericentric distance, $R_{peri}$, of a satellite is a more accurate predictor of its type of orbit than the more usual measure of eccentricity, $\eta$.
\end{itemize}
We note that our new model has a limited dependence on satellite mass, $M_{sat}$, which can be removed by using the specific energy $E$ and angular momentum $L$ instead. However, this has negligible effects on the accuracy of our model predictions, giving at most a 10\% difference for the largest satellites.

Our new model has implications for a wide range of problems in galaxy formation and evolution. As will be shown in future work, an implementation of our new model in the \SHA\ semi-analytic model of galaxy formation \citep{lagos_shark:_2018} has important implications for the contribution of satellite and central galaxies to the stellar mass function (SMF). Relative to the standard \SHA\ model, the numbers of satellites increase at all redshifts and all stellar masses; in contrast, the number of centrals at higher M$_{\star}$ is suppressed at $z$>0.5 because of the reduced merger rate, while it is enhanced at $z$=0 because super massive black holes are less massive and produce weaker feedback, leading to increased star formation rates and higher stellar masses. Further details will be presented in Proctor et al. (in preparation).

 We expect our new dynamical friction timescale to have an impact in a variety of areas, some of which we list below. 
 (i) Because our dynamical friction model affects the numbers of satellite and central galaxies, it is natural to expect changes in the predicted clustering of galaxies in mock galaxy catalogues,  which are important for comparison with galaxy surveys (e.g. \citealt{robotham_galaxy_2011}). 
 (ii) The new dynamical friction model presented here can be used to investigate halo-halo merger rates using extended Press-Schechter theory; cosmological simulations have shown that significant amount of dark matter in halos accumulate via mergers \citep{wright2020}, and hence an accurate understanding of dynamical friction for reliable predictions is required \citep{lacey_merger_1993,Parkinson2008}. (iii) We also expect our dynamical friction timescale to impact scaling relations of halos, particularly at the regime of groups and clusters, where the satellite population becomes increasingly important. An example of this is the HI-halo mass relation;  because HI in high halo mass systems resides predominantly within satellite galaxies, accurate lifetimes for satellites are essential if we are to produce strong
theoretical limits on the relation \citep{chauhan2020}.

Finally, we note that we have focused on the results of dark matter only simulations, but the
effect of baryons is non-negligible. Previous studies
\citep{boylan-kolchin_dynamical_2008,Dolag2009,Jiang2010} have found that baryons can reduce lifetimes of satellites/subhalos by $\approx 10$\% on average. Why exactly this occurs is interesting. The presence of baryons
tends to make subhalos hosting satellites more concentrated, and less susceptible to tidal stripping, which means that they can be exposed to the strong tides associated with the central galaxy within the host for longer - accelerating orbital decay. We hope to investigate further the
various factors that influence the merging timescale in the presence of baryons using hydrodynamical simulations in the future.

\section*{Acknowledgements}

We would like to thank Ainulnabilah B. Nasirudin for their assistance in preparing this manuscript. RP is supported by a University of Western Australia Scholarship, while PJE is
supported by a PDRA funded by the ARC Centre of Excellence in All-Sky Astrophysics in 3D 
(ASTRO 3D). Parts of this research was supported by the ARC Centre of Excellence ASTRO 3D through project number CE170100013. Part of this research was undertaken on Gadi, the NCI National Facility in Canberra, Australia, which is supported by the Australian commonwealth Government. 

\section*{Data availability}
The data used in this article were generated using the National Computing Infrastructure (NCI) high performance computing facility in Canberra, Australia. The derived data generated in this research will be shared on reasonable request to the corresponding author.




\bibliographystyle{mnras}
\bibliography{library} 

\begin{thebibliography}{}
\makeatletter
\relax
\def\mn@urlcharsother{\let\do\@makeother \do\$\do\&\do\#\do\^\do\_\do\%\do\~}
\def\mn@doi{\begingroup\mn@urlcharsother \@ifnextchar [ {\mn@doi@}
  {\mn@doi@[]}}
\def\mn@doi@[#1]#2{\def\@tempa{#1}\ifx\@tempa\@empty \href
  {http://dx.doi.org/#2} {doi:#2}\else \href {http://dx.doi.org/#2} {#1}\fi
  \endgroup}
\def\mn@eprint#1#2{\mn@eprint@#1:#2::\@nil}
\def\mn@eprint@arXiv#1{\href {http://arxiv.org/abs/#1} {{\tt arXiv:#1}}}
\def\mn@eprint@dblp#1{\href {http://dblp.uni-trier.de/rec/bibtex/#1.xml}
  {dblp:#1}}
\def\mn@eprint@#1:#2:#3:#4\@nil{\def\@tempa {#1}\def\@tempb {#2}\def\@tempc
  {#3}\ifx \@tempc \@empty \let \@tempc \@tempb \let \@tempb \@tempa \fi \ifx
  \@tempb \@empty \def\@tempb {arXiv}\fi \@ifundefined
  {mn@eprint@\@tempb}{\@tempb:\@tempc}{\expandafter \expandafter \csname
  mn@eprint@\@tempb\endcsname \expandafter{\@tempc}}}

\bibitem[\protect\citeauthoryear{Alves, Combes, Ferrara, Forveille  \&
  Shore}{Alves et~al.}{2016}]{planck_collaboration_planck_2015}
Alves J.,  Combes F.,  Ferrara A.,  Forveille T.,   Shore S.,  2016, \mn@doi [A
  {\&} A] {10.1051/0004-6361/201629543}, 594, E1

\bibitem[\protect\citeauthoryear{Baugh}{Baugh}{2006}]{baugh_primer_2006}
Baugh C.~M.,  2006, \mn@doi [Reports Prog. Phys.]
  {10.1088/0034-4885/69/12/R02}, 69, 3101

\bibitem[\protect\citeauthoryear{Benson}{Benson}{2010}]{benson_galaxy_2010}
Benson A.~J.,  2010, \mn@doi [Phys. Rep.] {10.1016/j.physrep.2010.06.001}, 495,
  33

\bibitem[\protect\citeauthoryear{Binney \& Tremaine}{Binney \&
  Tremaine}{1987}]{Binney1987}
Binney J.,  Tremaine S.,  1987, \mn@doi [Princet. NJ Princet. Univ. Press]
  {10.1063/1.3141945}

\bibitem[\protect\citeauthoryear{Blumenthal, Faber, Flores  \&
  Primack}{Blumenthal et~al.}{1986}]{Blumenthal1986}
Blumenthal G.~R.,  Faber S.~M.,  Flores R.,   Primack J.~R.,  1986, \mn@doi
  [ApJ] {10.1086/163867}, 301, 27

\bibitem[\protect\citeauthoryear{Boylan-Kolchin, Ma  \&
  Quataert}{Boylan-Kolchin et~al.}{2008}]{boylan-kolchin_dynamical_2008}
Boylan-Kolchin M.,  Ma C.~P.,   Quataert E.,  2008, \mn@doi [MNRAS]
  {10.1111/j.1365-2966.2007.12530.x}, 383, 93

\bibitem[\protect\citeauthoryear{Ca{\~{n}}as, Elahi, Welker, Lagos, Power,
  Dubois  \& Pichon}{Ca{\~{n}}as et~al.}{2019}]{Canas2019}
Ca{\~{n}}as R.,  Elahi P.~J.,  Welker C.,  Lagos C. d.~P.,  Power C.,  Dubois
  Y.,   Pichon C.,  2019, \mn@doi [MNRAS] {10.1093/mnras/sty2725}, 482, 2039

\bibitem[\protect\citeauthoryear{Chandrasekhar}{Chandrasekhar}{1943}]{Chandrasekhar1943}
Chandrasekhar S.,  1943, \mn@doi [Rev. Mod. Phys.] {10.1103/RevModPhys.15.1},
  15, a

\bibitem[\protect\citeauthoryear{{Chauhan}, {Lagos}, {Stevens}, {Obreschkow},
  {Power}  \& {Meyer}}{{Chauhan} et~al.}{2020}]{chauhan2020}
{Chauhan} G.,  {Lagos} C.~D.~P.,  {Stevens} A.~R.~H.,  {Obreschkow} D.,
  {Power} C.,   {Meyer} M.,  2020, arXiv e-prints, \href
  {https://arxiv.org/abs/2006.12102} {p. arXiv:2006.12102}

\bibitem[\protect\citeauthoryear{Cole, Lacey, Baugh  \& Frenk}{Cole
  et~al.}{2002}]{Shaun_Cole_et_al_2002}
Cole S.,  Lacey C.~G.,  Baugh C.~M.,   Frenk C.~S.,  2002, \mn@doi [MNRAS]
  {10.1046/j.1365-8711.2000.03879.x}, 319, 168

\bibitem[\protect\citeauthoryear{Colpi, Mayer  \& Governato}{Colpi
  et~al.}{1999}]{Colpi1999}
Colpi M.,  Mayer L.,   Governato F.,  1999, \mn@doi [ApJ] {10.1086/307952},
  525, 720

\bibitem[\protect\citeauthoryear{D'Onghia, Springel, Hernquist  \&
  Keres}{D'Onghia et~al.}{2010}]{DOnghia2010}
D'Onghia E.,  Springel V.,  Hernquist L.,   Keres D.,  2010, \mn@doi [ApJ]
  {10.1088/0004-637X/709/2/1138}, 709, 1138

\bibitem[\protect\citeauthoryear{Dayal \& Ferrara}{Dayal \&
  Ferrara}{2018}]{Dayal2018}
Dayal P.,  Ferrara A.,  2018, \mn@doi [Phys. Rep.]
  {10.1016/j.physrep.2018.10.002}, 780-782, 1

\bibitem[\protect\citeauthoryear{Deason, Wetzel  \& Garrison-Kimmel}{Deason
  et~al.}{2014}]{Deason2014}
Deason A.,  Wetzel A.,   Garrison-Kimmel S.,  2014, \mn@doi [ApJ]
  {10.1088/0004-637X/794/2/115}, 794

\bibitem[\protect\citeauthoryear{Dekel, Devor  \& Hetzroni}{Dekel
  et~al.}{2003}]{Dekel2003}
Dekel A.,  Devor J.,   Hetzroni G.,  2003, \mn@doi [MNRAS]
  {10.1046/j.1365-8711.2003.06432.x}, 341, 326

\bibitem[\protect\citeauthoryear{Dolag, Borgani, Murante  \& Springel}{Dolag
  et~al.}{2009}]{Dolag2009}
Dolag K.,  Borgani S.,  Murante G.,   Springel V.,  2009, \mn@doi [MNRAS]
  {10.1111/j.1365-2966.2009.15034.x}, 399, 497

\bibitem[\protect\citeauthoryear{Elahi, Thacker  \& Widrow}{Elahi
  et~al.}{2011}]{elahi_peaks_2011}
Elahi P.~J.,  Thacker R.~J.,   Widrow L.~M.,  2011, \mn@doi [MNRAS]
  {10.1111/j.1365-2966.2011.19485.x}, 418, 320

\bibitem[\protect\citeauthoryear{Elahi et~al.,}{Elahi
  et~al.}{2013}]{elahi_streams_2013}
Elahi P.~J.,  et~al., 2013, \mn@doi [MNRAS] {10.1093/mnras/stt825}, 433, 1537

\bibitem[\protect\citeauthoryear{Elahi, Welker, Power, Lagos, Robotham,
  Ca{\~{n}}as  \& Poulton}{Elahi et~al.}{2018}]{elahi_surfs:_2018}
Elahi P.~J.,  Welker C.,  Power C.,  Lagos C. d.~P.,  Robotham A.~S.,
  Ca{\~{n}}as R.,   Poulton R.,  2018, \mn@doi [MNRAS] {10.1093/mnras/sty061},
  475, 5338

\bibitem[\protect\citeauthoryear{Elahi, Ca{\~{n}}as, Poulton, Tobar, Willis,
  Lagos, Power  \& Robotham}{Elahi et~al.}{2019a}]{Elahi2019VELOCIraptor}
Elahi P.~J.,  Ca{\~{n}}as R.,  Poulton R.~J.,  Tobar R.~J.,  Willis J.~S.,
  Lagos C. D.~P.,  Power C.,   Robotham A.~S.,  2019a, \mn@doi [PASA]
  {10.1017/pasa.2019.12}, 36

\bibitem[\protect\citeauthoryear{Elahi, Poulton, Tobar, Canas, Lagos, Power  \&
  Robotham}{Elahi et~al.}{2019b}]{Elahi2019TreeFrog}
Elahi P.~J.,  Poulton R. J.~J.,  Tobar R.~J.,  Canas R.,  Lagos C. d.~P.,
  Power C.,   Robotham A. S.~G.,  2019b, \mn@doi [PASA] {10.1017/pasa.2019.18},
  36

\bibitem[\protect\citeauthoryear{Faber \& Jackson}{Faber \&
  Jackson}{1976}]{Faber1976}
Faber S.~M.,  Jackson R.~E.,  1976, \mn@doi [ApJ] {10.1086/154215}, 204, 668

\bibitem[\protect\citeauthoryear{Freeman}{Freeman}{1970}]{Freeman1970}
Freeman K.~C.,  1970, \mn@doi [ApJ] {10.1086/150474}, 160, 811

\bibitem[\protect\citeauthoryear{Gan, Kang, Hou  \& Chang}{Gan
  et~al.}{2010}]{gan_modelling_2010}
Gan J.~L.,  Kang X.,  Hou J.~L.,   Chang R.~X.,  2010, \mn@doi [Res. Astron.
  Astrophys.] {10.1088/1674-4527/10/12/005}, 10, 1242

\bibitem[\protect\citeauthoryear{Gnedin, Hernquist  \& Ostriker}{Gnedin
  et~al.}{1999}]{Gnedin1999}
Gnedin O.~Y.,  Hernquist L.,   Ostriker J.~P.,  1999, \mn@doi [ApJ]
  {10.1086/306910}, 514, 109

\bibitem[\protect\citeauthoryear{Hayashi, Navarro, Taylor, Stadel  \&
  Quinn}{Hayashi et~al.}{2003}]{Hayashi2003}
Hayashi E.,  Navarro J.~F.,  Taylor J.~E.,  Stadel J.,   Quinn T.,  2003,
  \mn@doi [ApJ] {10.1086/345788}, 584, 541

\bibitem[\protect\citeauthoryear{Jiang \& Binney}{Jiang \&
  Binney}{2000}]{Jiang2000}
Jiang I.-G.,  Binney J.,  2000, \mn@doi [MNRAS]
  {10.1046/j.1365-8711.2000.03311.x}, 314, 468

\bibitem[\protect\citeauthoryear{Jiang, Jing, Faltenbacher, Lin  \& Li}{Jiang
  et~al.}{2008}]{Jiang2008}
Jiang C.~Y.,  Jing Y.~P.,  Faltenbacher A.,  Lin W.~P.,   Li C.,  2008, \mn@doi
  [ApJ] {10.1086/526412}, 675, 1095

\bibitem[\protect\citeauthoryear{Jiang, Jing  \& Lin}{Jiang
  et~al.}{2010}]{Jiang2010}
Jiang C.~Y.,  Jing Y.~P.,   Lin W.~P.,  2010, \mn@doi [Astron. Astrophys.]
  {10.1051/0004-6361/200913257}, 510, A60

\bibitem[\protect\citeauthoryear{Jiang, Helly, Cole  \& Frenk}{Jiang
  et~al.}{2014}]{jiang_n-body_2014}
Jiang L.,  Helly J.~C.,  Cole S.,   Frenk C.~S.,  2014, \mn@doi [MNRAS]
  {10.1093/mnras/stu390}, 440, 2115

\bibitem[\protect\citeauthoryear{Knebe et~al.,}{Knebe
  et~al.}{2011}]{knebe_haloes_2011}
Knebe A.,  et~al., 2011, \mn@doi [MNRAS] {10.1111/j.1365-2966.2011.18858.x},
  415, 2293

\bibitem[\protect\citeauthoryear{Kravtsov, Gnedin  \& Klypin}{Kravtsov
  et~al.}{2004}]{Kravtsov2004a}
Kravtsov A.~V.,  Gnedin O.~Y.,   Klypin A.~A.,  2004, \mn@doi [ApJ]
  {10.1086/421322}, 609, 482

\bibitem[\protect\citeauthoryear{Lacey \& Cole}{Lacey \&
  Cole}{1993}]{lacey_merger_1993}
Lacey C.,  Cole S.,  1993, \mn@doi [MNRAS] {10.1093/mnras/262.3.627}, 262, 627

\bibitem[\protect\citeauthoryear{Lagos, Tobar, Robotham, Obreschkow, Mitchell,
  Power  \& Elahi}{Lagos et~al.}{2018}]{lagos_shark:_2018}
Lagos C. d.~P.,  Tobar R.~J.,  Robotham A.~S.,  Obreschkow D.,  Mitchell P.~D.,
   Power C.,   Elahi P.~J.,  2018, \mn@doi [MNRAS] {10.1093/mnras/sty2440},
  481, 3573

\bibitem[\protect\citeauthoryear{Lee et~al.,}{Lee
  et~al.}{2014}]{lee_sussing_2014}
Lee J.,  et~al., 2014, \mn@doi [MNRAS] {10.1093/mnras/stu2039}, 445, 4197

\bibitem[\protect\citeauthoryear{Miller, van~den Bosch, Green  \& Ogiya}{Miller
  et~al.}{2020}]{Miller2020}
Miller T.~B.,  van~den Bosch F.~C.,  Green S.~B.,   Ogiya G.,  2020,
  arXiv:2001.06489 [astro-ph.GA]

\bibitem[\protect\citeauthoryear{Mo, van~den Bosch  \& White}{Mo
  et~al.}{2010}]{Mo2010}
Mo H.,  van~den Bosch F.,   White S.,  2010, {Galaxy Formation and Evolution}.
Cambridge University Press, \mn@doi{10.1017/cbo9780511807244}

\bibitem[\protect\citeauthoryear{Muldrew et~al.,}{Muldrew
  et~al.}{2012}]{muldrew_measures_2012}
Muldrew S.~I.,  et~al., 2012, \mn@doi [MNRAS]
  {10.1111/j.1365-2966.2011.19922.x}, 419, 2670

\bibitem[\protect\citeauthoryear{Navarro, Frenk  \& White}{Navarro
  et~al.}{1995}]{Navarro1995}
Navarro J.~F.,  Frenk C.~S.,   White S. D.~M.,  1995, \mn@doi [MNRAS]
  {10.1093/mnras/275.1.56}, 275, 56

\bibitem[\protect\citeauthoryear{Navarro, Frenk  \& White}{Navarro
  et~al.}{1997}]{navarro_universal_1997}
Navarro J.~F.,  Frenk C.~S.,   White S. D.~M.,  1997, \mn@doi [ApJ]
  {10.1086/304888}, 490, 493

\bibitem[\protect\citeauthoryear{Ostriker, {Spitzer, Lyman}  \&
  Chevalier}{Ostriker et~al.}{1972}]{Ostriker1972}
Ostriker J.~P.,  {Spitzer, Lyman} J.,   Chevalier R.~A.,  1972, \mn@doi [ApJ]
  {10.1086/181018}, 176, L51

\bibitem[\protect\citeauthoryear{Parkinson, Cole  \& Helly}{Parkinson
  et~al.}{2008}]{Parkinson2008}
Parkinson H.,  Cole S.,   Helly J.,  2008, \mn@doi [MNRAS]
  {10.1111/j.1365-2966.2007.12517.x}, 383, 557

\bibitem[\protect\citeauthoryear{Poulton, Robotham, Power  \& Elahi}{Poulton
  et~al.}{2018}]{Poulton2018}
Poulton R.~J.,  Robotham A.~S.,  Power C.,   Elahi P.~J.,  2018, \mn@doi [PASA]
  {10.1017/pasa.2018.34}, 35

\bibitem[\protect\citeauthoryear{Poulton, Power, Robotham  \& Elahi}{Poulton
  et~al.}{2019}]{Poulton2019}
Poulton R. J.~J.,  Power C.,  Robotham A. S.~G.,   Elahi P.~J.,  2019, \mn@doi
  [MNRAS] {10.1093/mnras/stz3202}, 491, 3820

\bibitem[\protect\citeauthoryear{Robotham \& Obreschkow}{Robotham \&
  Obreschkow}{2015}]{Robotham2015}
Robotham A.~S.,  Obreschkow D.,  2015, \mn@doi [PASA] {10.1017/pasa.2015.33},
  32

\bibitem[\protect\citeauthoryear{Robotham et~al.,}{Robotham
  et~al.}{2011}]{robotham_galaxy_2011}
Robotham A.~S.,  et~al., 2011, \mn@doi [MNRAS]
  {10.1111/j.1365-2966.2011.19217.x}, 416, 2640

\bibitem[\protect\citeauthoryear{Rubin, Thonnard  \& {Ford, W. K.}}{Rubin
  et~al.}{1980}]{Rubin1980}
Rubin V.~C.,  Thonnard N.,   {Ford, W. K.} J.,  1980, \mn@doi [ApJ]
  {10.1086/158003}, 238, 471

\bibitem[\protect\citeauthoryear{Simha \& Cole}{Simha \&
  Cole}{2017}]{Simha2017}
Simha V.,  Cole S.,  2017, \mn@doi [MNRAS] {10.1093/MNRAS/STX1942}, 472, 1392

\bibitem[\protect\citeauthoryear{Somerville \& Dav{\'{e}}}{Somerville \&
  Dav{\'{e}}}{2015}]{somerville_physical_2015}
Somerville R.~S.,  Dav{\'{e}} R.,  2015, \mn@doi [Annu. Rev. Astron.
  Astrophys.] {10.1146/annurev-astro-082812-140951}, 53, 51

\bibitem[\protect\citeauthoryear{Srisawat et~al.,}{Srisawat
  et~al.}{2013}]{srisawat_sussing_2013}
Srisawat C.,  et~al., 2013, \mn@doi [MNRAS] {10.1093/mnras/stt1545}, 436, 150

\bibitem[\protect\citeauthoryear{Taffoni, Mayer, Colpi  \& Governato}{Taffoni
  et~al.}{2003}]{Taffoni2003}
Taffoni G.,  Mayer L.,  Colpi M.,   Governato F.,  2003, \mn@doi [MNRAS]
  {10.1046/j.1365-8711.2003.06395.x}, 341, 434

\bibitem[\protect\citeauthoryear{Taylor \& Babul}{Taylor \&
  Babul}{2004}]{Taylor2004}
Taylor J.~E.,  Babul A.,  2004, \mn@doi [MNRAS]
  {10.1111/j.1365-2966.2004.07395.x}, 348, 811

\bibitem[\protect\citeauthoryear{Toomre}{Toomre}{1977}]{Toomre1977}
Toomre A.,  1977, in Evol. Galaxies Stellar Popul.. p.~401

\bibitem[\protect\citeauthoryear{Velazquez \& White}{Velazquez \&
  White}{1999}]{Velazquez1999}
Velazquez H.,  White S. D.~M.,  1999, \mn@doi [MNRAS]
  {10.1046/j.1365-8711.1999.02354.x}, 304, 254

\bibitem[\protect\citeauthoryear{Wetzel \& White}{Wetzel \&
  White}{2010}]{Wetzel2010a}
Wetzel A.~R.,  White M.,  2010, \mn@doi [MNRAS]
  {10.1111/j.1365-2966.2009.16191.x}, 403, 1072

\bibitem[\protect\citeauthoryear{White \& Rees}{White \&
  Rees}{1978}]{White1978}
White S. D.~M.,  Rees M.~J.,  1978, \mn@doi [MNRAS] {10.1093/mnras/183.3.341},
  183, 341

\bibitem[\protect\citeauthoryear{{Wright}, {Lagos}, {Power}  \&
  {Mitchell}}{{Wright} et~al.}{2020}]{wright2020}
{Wright} R.~J.,  {Lagos} C. d.~P.,  {Power} C.,   {Mitchell} P.~D.,  2020,
  arXiv e-prints, \href {https://ui.adsabs.harvard.edu/abs/2020arXiv200600924W}
  {p. arXiv:2006.00924}

\bibitem[\protect\citeauthoryear{Zentner, Berlind, Bullock, Kravtsov  \&
  Wechsler}{Zentner et~al.}{2005}]{Zentner2005}
Zentner A.~R.,  Berlind A.~A.,  Bullock J.~S.,  Kravtsov A.~V.,   Wechsler
  R.~H.,  2005, \mn@doi [ApJ] {10.1086/428898}, 624, 505

\bibitem[\protect\citeauthoryear{van~den Bosch \& Ogiya}{van~den Bosch \&
  Ogiya}{2018}]{VandenBosch2018}
van~den Bosch F.~C.,  Ogiya G.,  2018, \mn@doi [MNRAS] {10.1093/MNRAS/STY084},
  475, 4066

\bibitem[\protect\citeauthoryear{van~den Bosch, Ogiya, Hahn  \&
  Burkert}{van~den Bosch et~al.}{2018}]{VandenBosch2018a}
van~den Bosch F.~C.,  Ogiya G.,  Hahn O.,   Burkert A.,  2018, \mn@doi [MNRAS]
  {10.1093/mnras/stx2956}, 474, 3043

\makeatother
\end{thebibliography}




\appendix

\section{Variation of pericentric radius with host virial radius} \label{app:CircVsR}

\begin{figure}
    \centering
    \includegraphics[width=0.49\textwidth]{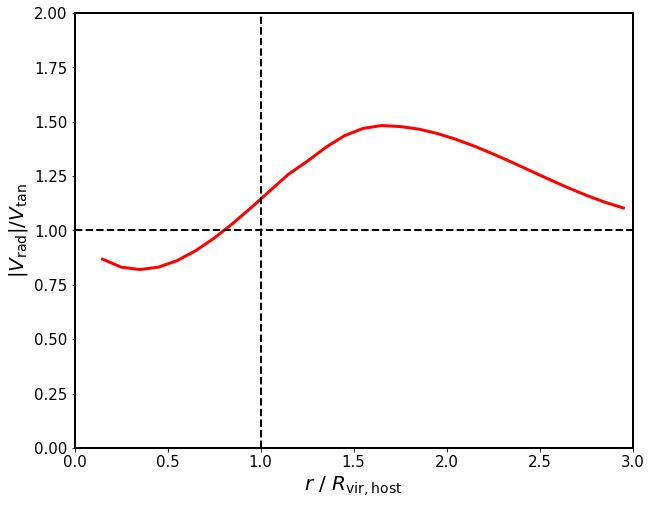}
    \caption{Variation of the ratio of the magnitude of radial velocity |$V_{rad}$| over the tangential velocity $V_{tan}$ as a function of $r$ / $R_{vir,host}$. The red line shows the median in each $r$ / $R_{vir,host}$ bin. The vertical dashed line shows when $r = R_{vir,host}$, while the horizontal dashed line shows where  |$V_{rad}$| = $V_{tan}$. \label{fig:VVsR}}
\end{figure}

The change in the magnitude of the radial velocity relative to the circular velocity, $V_{rad} / V_{circ}$, as a function of $r$ / $R_{vir,host}$, is shown in Figure \ref{fig:VVsR}. Objects on smoothly inspiralling orbits should have a constant $V_{rad} / V_{circ}$ with a value $\sim$1, which is not observed in Figure \ref{fig:VVsR}. The value for $V_{rad} / V_{circ}$ is greater than unity and varies with position, demonstrating that these objects tend to be on more radial orbits as they cross the region from 3 $R_{vir,host}$ to 1 $R_{vir,host}$. This tendency to
be on more radial orbits implies that the time to cross this region will be less than predicted assuming that the object is on a smoothly inspiralling orbit.

\section{Virial theorem} \label{app:virial}

\begin{figure}
    \centering
    \includegraphics[width=0.49\textwidth]{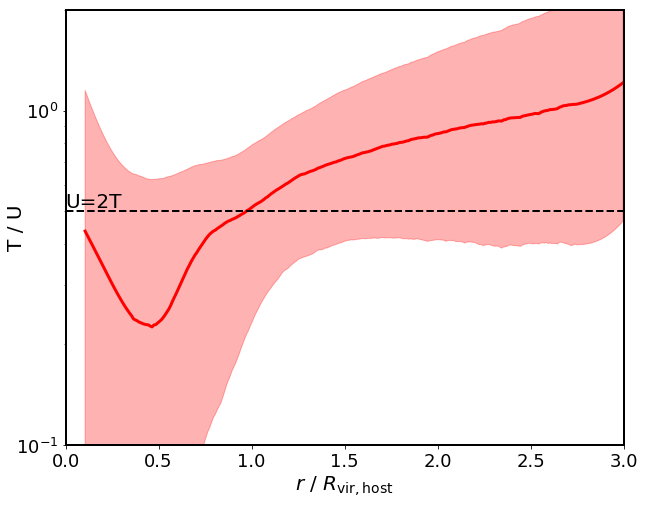}
    \caption{Variation of the ratio of satellites' kinetic to potential energy with $r$ / $R_{vir,host}$, where the red line shows the median in each $r$ / $R_{vir,host}$ bin and the shaded region shows the standard deviation. The black dashed line indicates the region where we would expect to find satellites that 
    follow the virial theorem (i.e. U=2T). \label{fig:EratioVsR}}
\end{figure}

Figure \ref{fig:EratioVsR} shows how the ratio of the kinetic to potential energy
of satellites changes with $r$/$R_{vir,host}$. The dashed line shows where the virial theorem is valid (U = 2T). Most satellites at large radii have U $>$ 2T and
this holds true until they pass within $R_{\rm vir,host}$, at which point the virial theorem holds for the satellite. However, as the satellite passes within $R_{\rm vir,host}$, it becomes more bound and so U/T decreases until 0.5$R_{\rm vir,host}$, at which point the small enclosed mass within the orbit of the satellite leads to U/T increasing.

\section{Orbit predictor} \label{app:orbpred}

\begin{figure}
    \centering
    \includegraphics[width=0.49\textwidth]{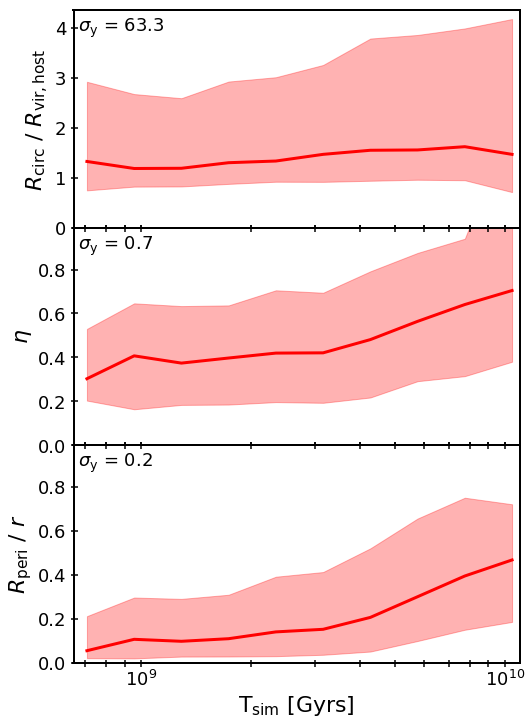}
    \caption{Variation of $R_{peri}$/$r$, $\eta$, and $R_{circ}$/$R_{vir,host}$ with simulation merger timescale T$_{sim}$, as computed at infall. The red line shows the median in each T$_{sim}$ bin and the shaded region shows the standard deviation. The text in the top left of each panel shows the standard deviation along the y-axis. \label{fig:OrbitPred}}
\end{figure}

The ideal properties for an orbit predictor are,
\begin{enumerate}[leftmargin=*]
\item its value should change with T$_{sim}$, demonstrating that it can predict whether an object is on a stable orbit with a long T$_{sim}$ or is on highly radial orbit with a short T$_{sim}$; and
\item there is limited scatter in its value because increased scatter leads to a larger uncertainty in T$_{model}$.
\end{enumerate}

We demonstrate some of the key orbit predictors explored in this work in Figure \ref{fig:OrbitPred}, where the top panel shows how the $R_{circ}/R_{vir,host}$ changes as a function of T$_{sim}$. $R_{circ}/R_{vir,host}$ is broadly insensitive to T$_{sim}$, which shows that it does not satisfy condition (i); it also has a large scatter in its value, which shows that it does not satisfy condition (ii). In contrast, $\eta$ (middle panel) shows a slight variation with T$_{sim}$, and so somewhat satisfies condition (i), while its reduced scatter satisfies condition (ii). However, we find that the best performing orbit predictor is $R_{peri} /r$, which shows the largest variation with T$_{sim}$ and the least amount of scatter.

\section{Satellite Selection Criteria} \label{app:Sel}

\begin{figure*}
    \centering
    \includegraphics[width=\textwidth]{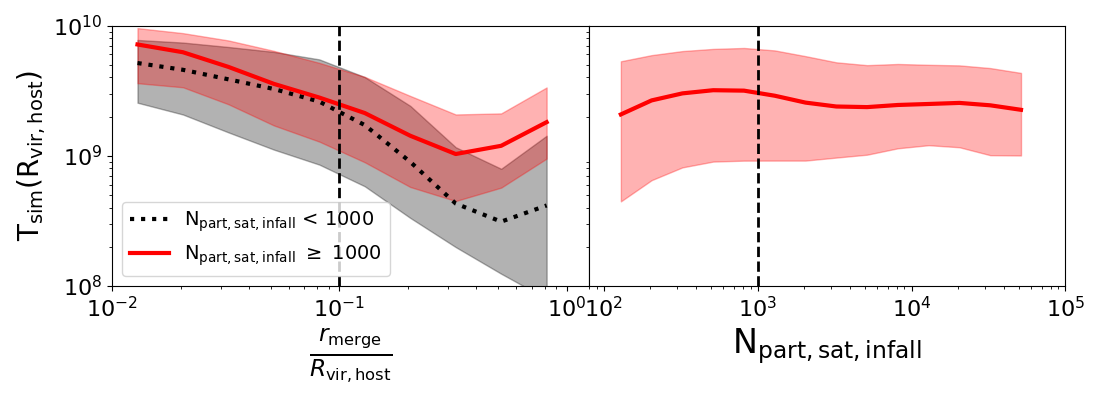}
    \caption[How T$_{\rm sim}$ calculated at infall varies with $\frac{r_{sat}}{R_{vir,host}}$ and N$_{part,sat,infall}$]{The left panel shows how the T$_{sim}$ calculated at infall (T$_{sim}$($R_{vir,host}$)) varies as a function of  where it merges within its host($\frac{r_{merge}}{R_{vir,host}}$) , where the solid red line shows the medians for satellites with N$_{part,sat,infall}$ $\geq$  1000 and the dotted black line shows for satellites with N$_{part,sat,infall}$ $<$  1000. The shaded regions shows the standard deviations for each selection. The right panel shows how T$_{sim}$($R_{vir,host}$) varies with   N$_{part,sat,infall}$, the solid red line shows the median and the shaded region shows the standard deviation. In both panels a selection of N$_{part,sat,infall}$ $\geq$  100 is also applied to remove unresolved satellites.  \label{fig:TsimSel}}
\end{figure*}

In this section, we evaluate the selection criteria for infalling satellite.  Figure \ref{fig:TsimSel} shows how T$_{sim}$ calculated at infall (T$_{sim}$($R_{vir,host}$)) changes with where the satellite merges within its host ($\frac{r_{merge}}{R_{vir,host}}$) and the number of particles the satellite had at infall (N$_{part,sat,infall}$) for all satellites that have N$_{part,sat,infall} >$ 100. From the figure, there is a slight bias towards longer T$_{sim}$($R_{vir,host}$) by selecting satellites that merge within 0.1 $R_{vir,host}$. This bias however, is only minor if selecting satellites that have N$_{part,sat,infall} >$ 1000. Furthermore, the right panel in Figure \ref{fig:TsimSel} shows that the selection of N$_{part,sat,infall} >$1000 cause little-to-no bias in T$_{sim}$(R$_{vir,host}$). 

\begin{figure}
    \centering
    \includegraphics[width=0.45\textwidth]{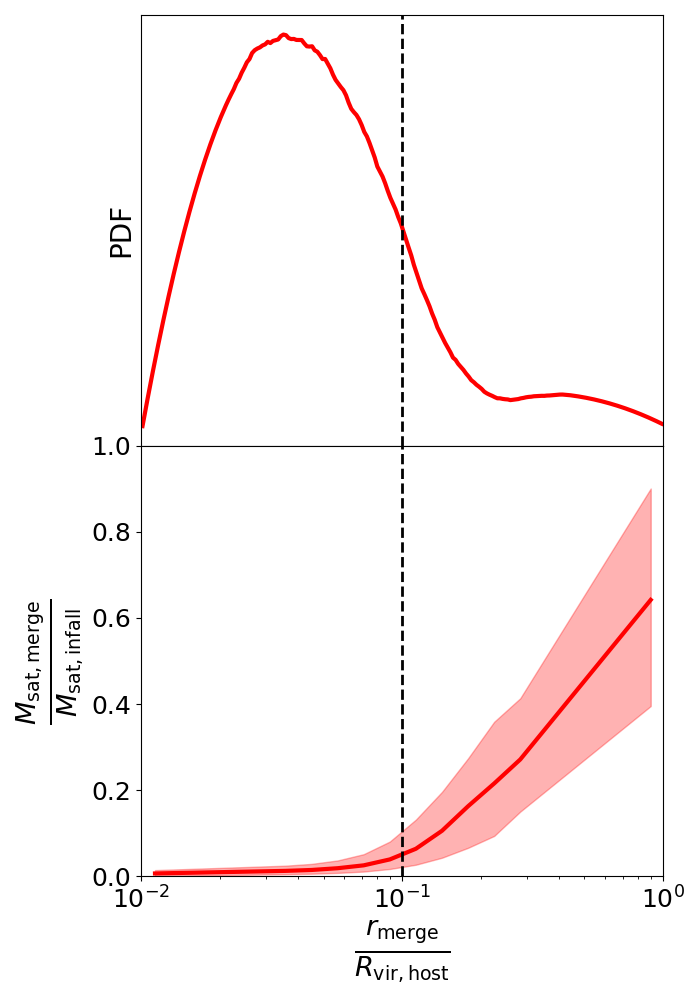}
    \caption[PDF of where satellites merge within their host and fractional mass loss of satellites as  a function of where they merge within their host]{The top panel shows a PDF of where satellites merge within $R_{vir,host}$ ($\frac{r_{merge}}{R_{vir,host}}$). The bottom panel shows the ratio of the mass that satellites merge with to the mass at infall as a function of $\frac{r_{merge}}{R_{vir,host}}$, where the red line is the median and the shaded region is the standard deviation. The vertical dashed line shows the merger selection criteria. \label{fig:rmergebias}}
\end{figure}

In Figure \ref{fig:rmergebias}, the top panel shows the PDF of where satellites with N$_{part,sat,infall} >$1000, merge within $R_{vir,host}$. From the plot it is evident that the selection of 0.1$R_{vir,host}$ captures the vast majority of satellites, with very few merging outside 0.1$R_{vir,host}$. We note that there is a small population of mergers outside of 0.1$R_{vir,host}$, which is most like due to satellite-satellite interaction. The bottom panel shows ratio of the mass the satellite merged with $M_{sat,infall}$ to the mass of the satellite at infall $M_{sat,infall}$ for satellites with N$_{part,sat,infall} >$1000 as a function of where they merge within $R_{vir,host}$. From the plot, satellites lose over 95\% of their mass when they merge within  0.1$R_{vir,host}$, meaning that these satellites have fully merged with their host.

\section{Convergence test} \label{app:conv}

\begin{figure*}
    \centering
    \includegraphics[width=\textwidth]{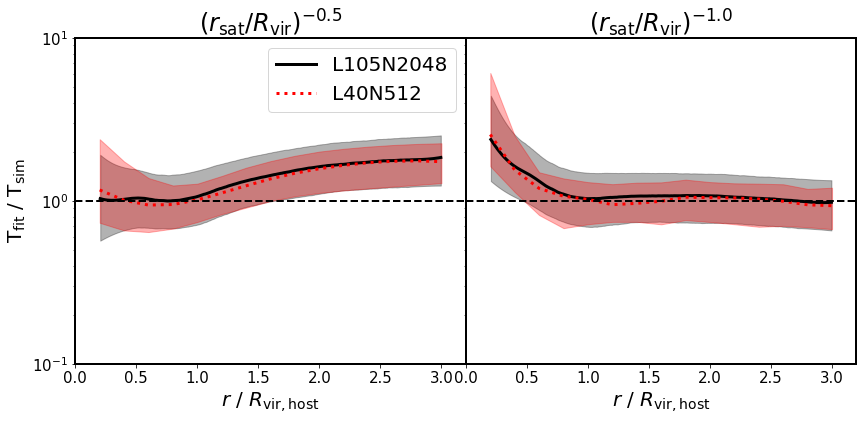}
    \caption[The prediction of the merger timescale presented in this work as a function of position relative to its host centre for a higher and a lower resolution simulations]{Plots showing the ratio of T$_{fit}$ to T$_{sim}$ for both the lower resolution 40 $h^{-1} {\rm Mpc}$ box with 512$^3$ particles (L40N512, red dotted line) and the higher resolution 105 $h^{-1} {\rm Mpc}$ box with 2048$^3$ particles (L105N2048, black solid line). For both,the lines are the medians and the shaded region shows the standard deviation. Each panel shows the different functional form, where the left panel shows Equation \ref{equ:Tnewin} and the right panel shows Equation \ref{equ:Tnewout}.    }
    \label{fig:tnewcomp}
\end{figure*}

To test the applicability of the new formula to other simulations, we apply it to a lower resolution 40 $h^{-1} {\rm Mpc}$ box with 512$^3$ particles from the Synthetic UniveRses For Surveys \citep{elahi_surfs:_2018} (SURFS) suite of simulations (with the same cosmology). We apply the same selection for infalling satellites by applying a mass cut of 1.73 $\times10^{10}$M$_{\odot}$\footnote{1000 particles at infall for the 105 $h^{-1} {\rm Mpc}$ box with 2048$^3$ particles, which corresponds to 419 particles in the 40 $h^{-1} {\rm Mpc}$ box with 512$^3$ particles.}, with all other selections the same. Figure \ref{fig:tnewcomp} shows the comparison of T$_{fit}$/T$_{sim}$ for the two simulations, where both simulations overlap at all $r/R_{vir,host}$ and have very similar scatter, with the higher resolution having slightly more scatter due to probing more types of satellites orbits.


\bsp	
\label{lastpage}
\end{document}